\documentclass[aps,twocolumn,showpacs,pra,groupedaddress,superscriptaddress,nofootinbib]{revtex4}
\usepackage{graphicx,amsmath,amssymb,amsbsy,subfigure,hyperref,bbm,times,txfonts}
\hypersetup{
   colorlinks=true,
   linkcolor=blue,
}
\graphicspath{{figure/}}
\begin{document}

\newcommand{\ket}[1]{|#1\rangle}
\newcommand{\bra}[1]{\langle#1|}
\newcommand{\ketbra}[1]{| #1\rangle\!\langle #1 |}
\newcommand{\kebra}[2]{| #1\rangle\!\langle #2 |}
\newcommand{\id}{\mathbbm{1}}
\newcommand{\ohm}{\Omega_{\rm CQ}}
\newcommand{\rhobd}{\rho^{\vec{c}}_{AB}}
\providecommand{\tr}[1]{\text{tr}\left[#1\right]}
\providecommand{\tra}[1]{\text{tr}_A\left[#1\right]}
\providecommand{\trb}[1]{\text{tr}_B\left[#1\right]}
\providecommand{\abs}[1]{\left|#1\right|}
\providecommand{\sprod}[2]{\langle#1|#2\rangle}
\providecommand{\expect}[2]{\bra{#2} #1 \ket{#2}}


\title{Protecting quantum resources via frequency modulation of qubits in leaky cavities}

\author{Ali Mortezapour}
\email{mortezapour@guilan.ac.ir}
\affiliation{Department of Physics, University of Guilan, P. O. Box 41335--1914, Rasht, Iran}
\author{Rosario Lo Franco}
\email{rosario.lofranco@unipa.it}
\affiliation{Dipartimento di Energia, Ingegneria dell'Informazione e Modelli Matematici, Universit\`{a} di Palermo, Viale delle Scienze, Edificio 9, 90128 Palermo, Italy}
\affiliation{Dipartimento di Fisica e Chimica, Universit\`a di Palermo, via Archirafi 36, 90123 Palermo, Italy}

\begin{abstract}
Finding strategies to preserve quantum resources in open systems is nowadays a main requirement for reliable quantum-enhanced technologies. We address this issue by considering structured cavities embedding qubits driven by a control technique known as frequency modulation. We first study a single qubit in a lossy cavity to determine optimal modulation parameters and qubit-cavity coupling regime allowing a gain of four orders of magnitude concerning coherence lifetimes. We relate this behavior to the inhibition of the qubit effective decay rate rather than to stronger memory effects (non-Markovianity) of the system. We then exploit these findings in a system of noninteracting qubits embedded in separated cavities to gain basic information about scalability of the procedure. We show that the determined modulation parameters enable lifetimes of quantum resources, such as entanglement, discord and coherence, three orders of magnitude longer than their natural (uncontrolled) decay times. We discuss the feasibility of the system within the circuit-QED scenario, typically employed in the current quantum computer prototypes. These results provide new insights towards efficient experimental strategies against decoherence.  
\end{abstract}

\date{\today}


\maketitle

\section*{Introduction}
Quantum coherence, stemming from the superposition principle, is the key concept that distinguishes quantum mechanics from classical mechanics. It is well-known that quantum coherence among different bodies of a multipartite systems is the basic ingredient for the formation of quantum correlations \cite{ref1,ref2,ref3,ref4,ref5,ref6,ref7,ref8,ref9,ref10}. Among these, entanglement represents the part due to non-separability of states, utilized in many paradigmatic quantum information processes such as teleportation \cite{ref11,pirandolaReview,lofrancoIndistResource}, quantum error correction \cite{ref12,ref13}, quantum key distribution \cite{ref14} and quantum dense coding \cite{ref15}. Useful quantum correlations beyond entanglement, occurring even for separable states, have been identified by the so-called discord \cite{ref18,ref19}, which is in turn employed for specific quantum information tasks \cite{ref16,ref17,ModiRMP}. As a very fundamental trait, quantum coherence itself can be quantified \cite{ref20, ref25,ref26,ref27,ref28,ref24} and exploited as a resource in various quantum protocols \cite{ref20,ref21,ref22,ref23,ref24}. One of the cutting-edge fields of application of these quantum features is, for instance, quantum metrology, aiming at achieving high precision measurements characterized by the quantum Fisher information \cite{Metro1,Metro2,Metro3,Metro4,Metro5,Metro6,Metro7,Metro8,
Metro9,Metro10,Metro11,Metro12}, with recent developments in open quantum systems \cite{Metro8,Metro13,Metro10,Metro15,Metro16,Metro17}. 

Open quantum systems undergo decoherence due to the interaction with the surrounding environment, which usually destroys coherence and correlations thus limiting their practical use in quantum information \cite{ref39} and metrology \cite{Metro15,Metro18,Metro5,Metro20,Metro21}. 
The dynamics of open quantum systems can be classified in Markovian (memoryless) and non-Markovian (memory-keeping) regimes \cite{ref30,ref31}. Markovian regime is typically associated to the lack of memory effects, a situation where the information irreversibly flows from the system to the environment. Differently, non-Markovian regime implies that the past history of the system affects the present one: such a memory effect results in an information feedback from the environment to the system \cite{ref30,ref31,ref32,ref34,ref35,ref36,ref37,ref38,ref39,ref40,ref41,PhysRevA.95.052126}. In this sense, non-Markovianity has been quantified by a variety of measures \cite{ref41,ref44,ref45,ref46,ref47,ref48,ref49,ref50,ref43,ref52,ref53} and regarded as a precious resource for certain applications \cite{ref41,ref42,ref43}. Although memory effects permit coherence, entanglement and discord of noninteracting subsystems to partially revive after disappearing during the time evolution \cite{ref31,ref39,ref46, ref54,ref55,ref56,ref57}, these revivals eventually decay. Decoherence remains one of the main drawbacks to overcome towards the implementation of quantum-enhanced technologies, which require qubits with long-lived quantum features. To this purpose, many strategies have been proposed in order to harnessing and protecting quantum resources in systems of qubits under different environmental conditions \cite{lofrancoreview,ref58,ref59,ref60,ref61,ref62,ref63,ref64,lofrancospinecho,
ref65,ref66,ref67,ref68,ref69,darrigo2012AOP,orieux2015,ref70,ref71,ref72,
LoFrancoNatCom,ref73,ref74,ref75,ref76,ref77,
ref78,lofranco2012PRA,ref79,ref80,ref81,ref82,
darrigo2013hidden,darrigo2014IJQI,ref83,ref84,ref85,
ref86,ref87,ref88,ref89,ref90,Ali2017}.

In this paper, we address this issue by using qubit frequency modulation as control technique. Generally, a quantum system is frequency modulated when its energy levels are shifted by an external driving. Frequency modulation in an atomic qubit can be performed by applying an external off-resonant field \cite{ref91,ref92,ref93}. Moreover, the most recent experimental progress in fabrication and control of quantum circuit-QED devices enables frequency modulation in superconducting Josephson qubits (artificial atoms) \cite{ref93,ref94,ref95,ref96,ref97}, which are the preferred building blocks in current quantum computer prototypes \cite{NatNews2017b}. 
It has been reported that external control of the qubit frequency can induce sidebands transitions \cite{ref98,ref99}, modify its fluorescence spectrum \cite{ref100} and population dynamics \cite{ref101,ref102,ref103,ref104,macoveiPRA}, as well as amplify non-Markovianity \cite{poggi2017}. 
Motivated by these considerations, here we aim at providing a thorough quantitative analysis of the role frequency modulation can play in prolonging lifetimes of desired quantum resources. Differently from previous studies, we mainly focus on determining the values of parameters of the driving field and of the qubit-cavity coupling which can efficiently maintain coherence and correlations for times orders of magnitude longer than their natural lifetimes in absence of external control. We first consider a single frequency-modulated qubit inside a structured leaky cavity, which allows us to determine the desired values of parameters. We also analyze how the choice of these parameters plays a role in preserving quantum Fisher information so to enhance the precision of phase estimation in the system during the evolution. The reliability of the theoretical results is checked by taking into account  experimental parameters typical of the circuit-QED scenario for a transmon qubit in a coplanar resonator.
We then extend the analysis to a system of two noninteracting qubits in separated cavities for establishing at which extent entanglement, discord and coherence can be shielded from decay thanks to individual control of the qubit frequencies. This novel investigation of simple systems constitutes the essential step for acquiring useful insights which can be straightforwardly generalized to a system of many separated qubits for scalability. 

The paper is organized as follows. In Sec.~\ref{secBell} we describe the evolution of the frequency-modulated open qubit, giving the dynamics of coherence and quantum Fisher information (QFI). In Sec.~\ref{secDM} we discuss the dynamics of quantum resources in the two-qubit system for some classes of initially correlated states. Finally, in Sec.~\ref{secU} we summarize the main conclusions.

\section*{Results}

\section{Single-qubit system}\label{secBell}

\begin{figure}[t!]
\begin{center}
\includegraphics[width=0.4\textwidth]{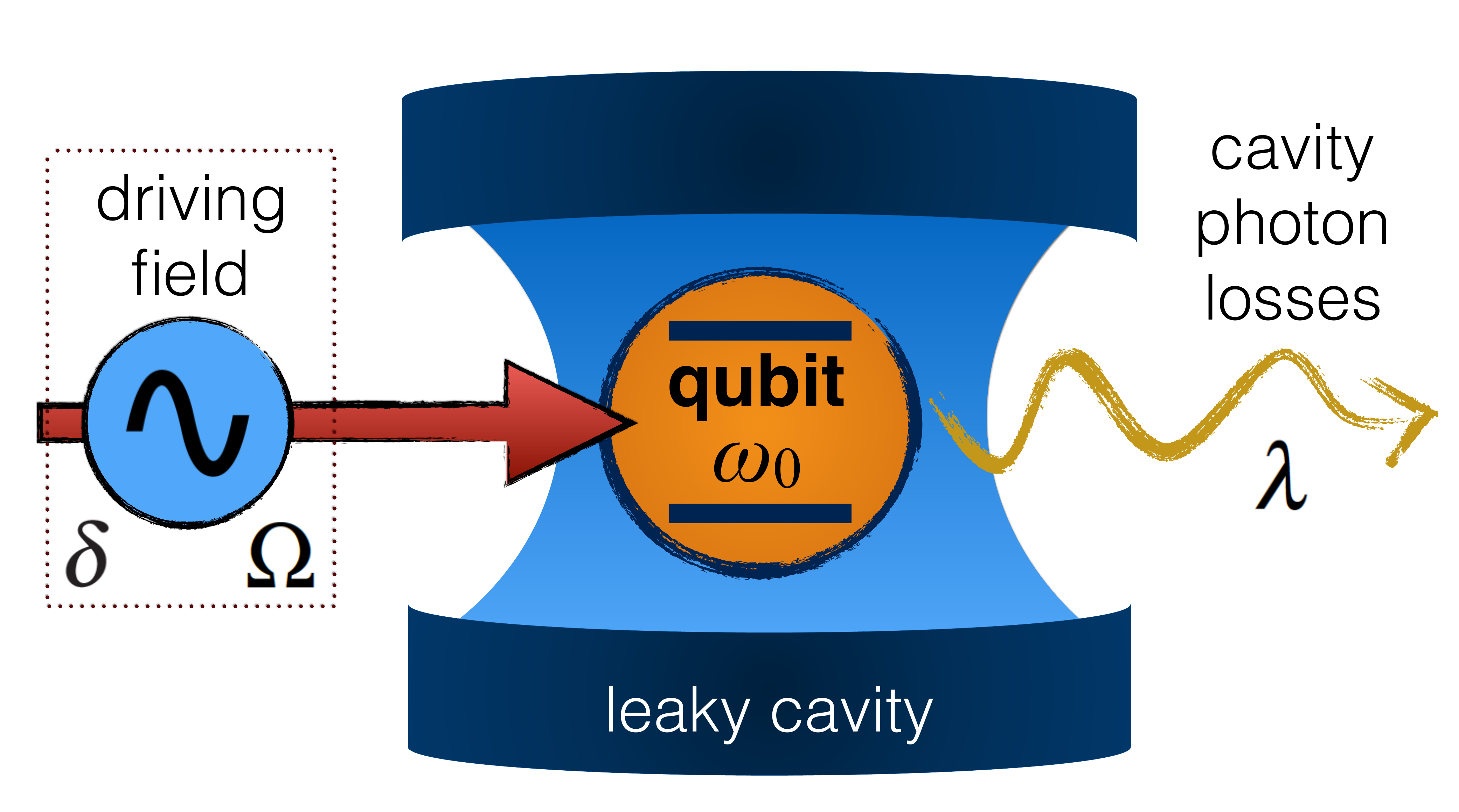}
\end{center}
\caption{Sketch of the single-qubit system. A qubit with transition frequency $\omega_0$ is embedded in a high-$Q$ leaky cavity, whose photon losses are characterized by the spectral width $\lambda\propto 1/Q$ of the coupling, assuming a Lorentzian spectrum for the cavity modes. The transition frequency of the qubit is modulated sinusoidally by an external driving field with a modulation amplitude $\delta$ and a modulation frequency $\Omega$.}
\label{AR1}
\end{figure}

The system consists of a qubit (two-level system) coupled to a zero-temperature reservoir formed by the quantized modes of a high-$Q$ cavity. The excited and ground states of the qubit are labeled by $\left| e \right\rangle $ and $\left| g \right\rangle$, respectively. It is assumed that the transition frequency of the qubit $\omega _{0} $ is modulated sinusoidally by an external driving field, as depicted in Fig.~\ref{AR1}. The Hamiltonian of the system is
\begin{equation} 
\label{eq:1} 
\hat{H}=\hat{H}_\mathrm{q} +\hat{H}_\mathrm{r} +\hat{H}_\mathrm{in}, 
\end{equation}
where $\hat{H}_\mathrm{q} $ is the qubit Hamiltonian ($\hbar =1$)
\begin{equation} 
\label{eq:2} 
\hat{H}_\mathrm{q} =\frac{1}{2} [\omega _{0} +\delta \cos (\Omega t)]\hat{\sigma }_{z} , 
\end{equation} 
with $\delta $ and $\Omega$ denoting the modulation amplitude and frequency, respectively, and $\hat{\sigma }_{z} $ the Pauli operator along the z-direction. $\hat{H}_\mathrm{r} =\sum _{k}\omega _{k} \hat{a}_{k}^{\dagger} \hat{a}_{k} $ is the reservoir Hamiltonian with $\hat{a}_{k}^{\dagger}$ ($\hat{a}_{k}$) being the creation (annihilation) operator of the cavity field mode $k$. $\hat{H}_\mathrm{in} $ describes the interaction between the qubit and the cavity modes which, in the dipole and rotating-wave approximation, can be written as
\begin{equation} 
\label{eq:3} 
\hat{H}_\mathrm{in} =\sum _{k}(g_{k} \hat{\sigma }_{+} \hat{a}_{k} +g_{k}^{*} \hat{a}_{k}^{\dagger} \hat{\sigma }_{-}),  
\end{equation}
where $g_{k} $ is the coupling constant between the qubit and mode $k$ while $\hat{\sigma }_{+}$ ($\hat{\sigma }_{-}$) represents the raising (lowering) operator for the qubit.
We point out that, in general, when a nonstationary mode (in this case, the qubit mode) is coupled to the environment, the system-reservoir couplings $g_k$ also become time dependent \cite{diogoPRA,celeriJPB}. However, here we focus on an experimentally feasible system where $\delta/\omega_0\ll 1$ and $\Omega/\omega_0\ll 1$ (see subsection~\ref{expcont}), which guarantee the so-called adiabatic regime: under such condition, the qubit-reservoir interaction is still modelled by the Jaynes-Cummings interaction, with the couplings $g_k$ independent of time, despite the nonstationary qubit mode \cite{celeriJPB,PhysRevA.57.4784}. Moreover, the qubit dimension is taken to be much smaller than the wavelength of the emitted photon so that the dipole approximation remains valid and, since $\omega_0 \pm \delta \approx \omega_0$, also the rotating wave approximation holds along our analysis \cite{poggi2017,ref102}.  

Moving to a non-uniformly rotating frame (interaction picture) by the unitary transformation
\begin{equation} 
\label{eq:4}
\hat{U}=\exp \left[-i\left\{\sum _{k}\omega _{k} \hat{a}_{k}^{\dagger} \hat{a}_{k} + [\omega _{0} +(\delta /\Omega )\sin \Omega t]\hat{\sigma }_{z} \right\}\right], 
\end{equation} 
the effective Hamiltonian $\hat{H}_\mathrm{eff} =\hat{U}^{\dagger} \hat{H}\hat{U}+(\partial \hat{U}^{\dagger} /\partial t)\hat{U}$ is
\begin{eqnarray}\label{eq:5}
\hat{H}_\mathrm{eff} &=& \sum _{k}g_{k} \hat{\sigma }_{+} \hat{a}_{k} e^{-i(\omega _{k} -\omega _{0} )t} e^{i(\delta /\Omega )\sin \Omega t} \nonumber \\
&+& \sum _{k} g_{k}^{*} \hat{a}_{k}^{ \dagger } \hat{\sigma }_{-} e^{i(\omega _{k} -\omega _{0} )t} e^{-i(\delta /\Omega )\sin \Omega t}\ .
\end{eqnarray} 
We notice that $\hat{H}_\mathrm{eff}$ shows how the system behaves as if the same frequency modulation was applied to each qubit-cavity coupling constant $g_k$. This is relevant to our purpose since, under this condition, it is known that decoherence can be softened \cite{agarwalPRA,macoveiPRA}. Making use of Jacobi-Anger expansion, the exponential factors in Eq.~(\ref{eq:5}) can be written as
\begin{equation} 
\label{eq:6}
e^{\pm i(\delta /\Omega )\sin (\Omega t)} =J_{0} \left(\frac{\delta }{\Omega } \right)+2\sum _{n=1}^{\infty }(\pm i)^{n} J_{n} \left(\frac{\delta }{\Omega } \right) \cos (n\Omega t), 
\end{equation} 
where $J_{n} \left(\frac{\delta }{\Omega }\right)$ is the $n$-th Bessel function of the first kind. 

We assume the qubit is initially in a coherent superposition $\alpha {\left| e \right\rangle} +\beta {\left| g \right\rangle} $ and the reservoir modes in the vacuum state ${\left| 0 \right\rangle} $, so that the overall initial state is
\begin{equation} 
\label{eq:7} 
\left| \Psi (0) \right\rangle =(\alpha \left| e \right\rangle +\beta \left| g \right\rangle) \left| 0 \right\rangle. 
\end{equation} 
Hence, at any later time $t$ the quantum state of the whole system can be written as
\begin{equation}  
\label{eq:8}
{\left| \Psi (t) \right\rangle} =\alpha C_{e} (t){\left| e \right\rangle} {\left| 0 \right\rangle} +\beta {\left| g \right\rangle} {\left| 0 \right\rangle} +\sum _{k}C_{g,k} (t){\left| g \right\rangle} {\left| 1_{k}  \right\rangle}  , 
\end{equation} 
where ${\left| 1_{k}  \right\rangle}$ is the cavity state with a single photon in mode $k$ and $C_{g,k} (t)$ is its probability amplitude. Using the time-dependent 
Schr{\"{o}}dinger equation, the differential equations for the probability amplitudes $C_{e} (t)$ and $C_{g,k} (t)$ are, respectively,
\begin{equation}  
\label{eq:9}
\dot{C}_{e} (t)=-i e^{i(\delta /\Omega )\sin (\Omega t)}\sum _{k}g_{k} e^{-i(\omega _{k} -\omega _{0} )t} C_{g,k} (t) ,                            
\end{equation}
and
\begin{equation} 
\label{eq:10}
\dot{C}_{g,k} (t)=-i e^{-i(\delta /\Omega )\sin (\Omega t)} g_{k}^{*} e^{i(\omega _{k} -\omega _{0} )t} C_{e} (t). 
\end{equation} 
Solving Eq.~(\ref{eq:10}) formally and substituting the solution into Eq.~(\ref{eq:9}), one obtains
\begin{equation} 
\label{eq:11} 
\dot{C}_{e} (t)+\int _{0}^{t}dt'  G(t,t' )C_{e} (t')=0, 
\end{equation} 
where the kernel $G(t,t')$, that is the correlation function including the memory effects, is
\begin{equation} 
\label{eq:12} 
G(t,t')=e^{i(\delta /\Omega )[\sin (\Omega t)-\sin(\Omega t')]}\sum _{k}\left|g_{k} \right|^{2} e^{-i(\omega _{k} -\omega _{0} )(t-t')}  . 
\end{equation} 
In the continuous limit, the kernel above becomes 
\begin{equation} 
\label{eq:13}
G(t,t' )=e^{i(\delta /\Omega )[\sin (\Omega t)-\sin(\Omega t')]}
 \int _{0}^{\infty }J(\omega)e^{-i(\omega-\omega_{0} )(t-t')}d\omega, 
\end{equation}
where $J(\omega)=\sum_k |g_k|^2\delta(\omega-\omega_k)$ is, as usual, the spectral density of the reservoir (cavity) modes \cite{ref30,poggi2017}. We choose a Lorentzian spectral density, which is typical of a structured cavity \cite{ref30}, whose form is
\begin{equation} 
\label{eq:14} 
J(\omega)=\frac{1}{2\pi } \frac{\gamma \lambda ^{2} }{(\omega_{0} -\omega)^{2} +\lambda ^{2} }, 
\end{equation} 
where $\lambda$ indicates the spectral width of the coupling and is related to the reservoir correlation time $\tau _{r} $ via $\tau_\mathrm{r} =\lambda ^{-1} $. On the other hand, $\gamma $ represents the decay rate of the excited state of the qubit in the Markovian limit of flat spectrum (i.e., the spontaneous emission decay rate) and it is linked to the qubit relaxation time $\tau_\mathrm{q}$ by $\tau_\mathrm{q} = \gamma ^{-1}$ \cite{ref30}. Qubit-cavity weak coupling
occurs for $\lambda>\gamma$ ($\tau_\mathrm{r}<\tau_\mathrm{q}$); the opposite condition $\lambda<\gamma$ ($\tau_\mathrm{r}>\tau_\mathrm{q}$) thus identifies strong coupling. The larger the cavity quality factor, the smaller the spectral width $\lambda$. 

With such a spectral density, the kernel of Eq.~(\ref{eq:13}) becomes
\begin{equation} 
\label{eq:15} 
G(t,t' )=\frac{\gamma \lambda }{2} e^{-\lambda (t-t')} 
e^{i(\delta /\Omega )[ \sin (\Omega t)-\sin (\Omega t')]}. 
\end{equation}
Substituting it into Eq.~(\ref{eq:11}), one gets   
\begin{equation}\label{eq:16}
\dot{C}_{e} (t)+\frac{\gamma \lambda}{2} e^{i(\delta/\Omega )
\sin (\Omega t)}  \int _{0}^{t}dt' e^{-i(\delta /\Omega )
\sin (\Omega t')}e^{-\lambda (t-t' )} C_{e} (t')=0.
\end{equation} 
Calculating $C_{e} (t)$ from this equation, the reduced density matrix of the qubit $\rho_\mathrm{q} (t)$ in the basis $\{\left| e \right\rangle, \left| g \right\rangle\}$ is given by                   
\begin{equation}
\label{eq:17}
\rho_\mathrm{q} (t)=
\left(
\begin{array}{cc}
{\left|\alpha \right|^{2} \left|C_{e} (t)\right|^{2} } & {\alpha \beta ^{*} C_{e} (t)} \\
{\alpha ^{*} \beta C_{e}^{*} (t)} & {1-\left|\alpha \right|^{2} \left|C_{e} (t)\right|^{2} } \\
\end{array}
\right).
\end{equation}
Taking the derivative of Eq.~(\ref{eq:16}) with respect to time, we have
\begin{eqnarray}
\label{eq:18}
\dot{\rho }_\mathrm{q} (t)&=&-i\frac{\Omega (t)}{2} [\sigma _{+} \sigma _{-} ,\rho_\mathrm{q} (t)] + \frac{\Gamma (t)}{2} (2\sigma _{-} \rho_\mathrm{q} (t)\sigma _{+}\nonumber\\
&&-\sigma _{+} \sigma _{-} \rho_\mathrm{q} (t)-\rho_\mathrm{q} (t)\sigma _{+} \sigma _{-}),
\end{eqnarray} 
where $\Omega (t)=-2\mathrm{Im}\left[\frac{\dot{C}_{e} (t)}{C_{e} (t)} \right]$ plays the role of a time-dependent Lamb shift and $\Gamma (t)=-2\mathrm{Re}\left[\frac{\dot{C}_{e} (t)}{C_{e} (t)} \right]$ can be interpreted as a time-dependent decay rate \cite{ref30}. 

The evolved density matrix of the qubit $\rho_\mathrm{q} (t)$ shall be utilized for obtaining the dynamics of the quantum properties of interest, such as coherence, quantum Fisher information and non-Markovianity. Before displaying the dynamics of these quantities, we recall their definitions and give their time-dependent expressions in the following.

\subsection{Coherence} 
Coherence of a quantum state characterizes the property of superposition among the basis states of the system and can be regarded as a basis-dependent resource by itself \cite{ref24}. Among the various bona-fide measures of coherence, we choose here the so-called $l_1$-norm defined as \cite{ref21}
\begin{equation} \label{eq:19} 
\zeta =\min\limits_{\rho_\mathrm{inc} \in \mathcal{I}}
||\rho - \rho_\mathrm{inc}||_{l_1} = \sum_{i\ne j}\left|\rho _{ij}\right|,
\end{equation}
where $\rho$ is the density matrix of an arbitrary quantum state and the minimum is taken over the set $\mathcal{I}$ of incoherent states $\rho_\mathrm{inc}$. Such a measure, depending only on the off-diagonal elements $\rho _{ij}$ ($i\ne j$) of the quantum state, is clearly related to the fundamental property of quantum interference. 

For $\rho_\mathrm{q} (t)$ of Eq.~(\ref{eq:18}) and assuming $\alpha =\beta =1/\sqrt{2}$ in the initial state of Eq.~(\ref{eq:7}), we have $\zeta(0)=1$ (maximum initial coherence) and a time-dependent qubit coherence $\zeta(t)=\left|C_{e} (t)\right|$.

\subsection{Quantum Fisher information} 
Quantum metrology exploits quantum-mechanical effects to reach high precision measurements. In a typical metrological procedure, one first encodes the parameter of interest $\phi$ on a probe state $\rho_\mathrm{in}$ by means of a unitary process $U_{\phi}$. Thus, the output state is $\rho_{\phi}= U_{\phi} \rho_{in} U_{\phi}^{\dagger}$. The output state $\rho_{\phi}$ is then measured by a set of positive operator valued measurements and the value of $\phi$ finally estimated from the outcomes. It is known that the precision in estimating $\phi$ is limited by the quantum Cramer-Rao bound inequality \cite{Metro22,Metro23}
\begin{equation} 
\label{eq:22}
\delta \phi \geq 1/\sqrt{F_{\phi}},
\end{equation} 
where $\delta \phi$ is the standard deviation associated to the variable $\phi$ and $F_{\phi}$ is the quantum Fisher information (QFI) defined as $F_{\phi}=\mathrm{Tr}(\rho_{\phi}L^2)$, with $L$ being the so-called symmetric logarithmic derivative determined by $\partial_{\phi}\rho_{\phi}=(L\rho(\phi)+\rho(\phi) L)/2$ with $\partial_{\phi}=\partial / \partial \phi$ \cite{Metro22,Metro23}. From the spectral decomposition of the $\phi$-dependent density matrix $\rho_{\phi}=\sum_{m} p_{m} \left| \psi_m \right\rangle \left\langle \psi_{m}\right|$, where $p_m$ and $\ket{\psi_m}$ are, respectively, eigenvalues and eigenstates, the QFI is known to have the analytical expression \cite{Metro6} 
\begin{equation} \label{eq:24}
F_{\phi}=\sum _{m,n} \frac{2}{p_m + p_n}
|\bra{\psi_m}\partial_{\phi}\rho_{\phi}\ket{\psi_n}|^2.
\end{equation}  

For our dynamical system, we consider a phase-estimation problem where 
$U_\phi\equiv \ket{g}\bra{g} + e^{i\phi}\ket{e}\bra{e}$ acts on the initial maximally coherent state of the qubit $\ket{\psi_+}=(\ket{e}+\ket{g})/\sqrt{2}$ \cite{Metro1} and successively let the system evolve under the dissipative noise and frequency modulation. The initial overall qubit-cavity state is therefore $(U_\phi \ket{\psi_+})\ket{0}$ and the evolved reduced density matrix of the qubit $\rho_{\mathrm{q},\phi}(t)$ has the form of Eq.~(\ref{eq:17}) with a $\phi$-dependence in the off-diagonal elements. Using Eq.~(\ref{eq:24}) we get $F_{\phi}(t)=|C_{e} (t)|^{2}$, that is the coherence squared. The associated minimum estimation error is therefore 
$\delta\phi_\mathrm{min}(t)=1/\sqrt{F_{\phi}(t)}= 1/|C_{e} (t)|$.

\begin{figure}[t!]
\begin{center}
\includegraphics[scale=.62]{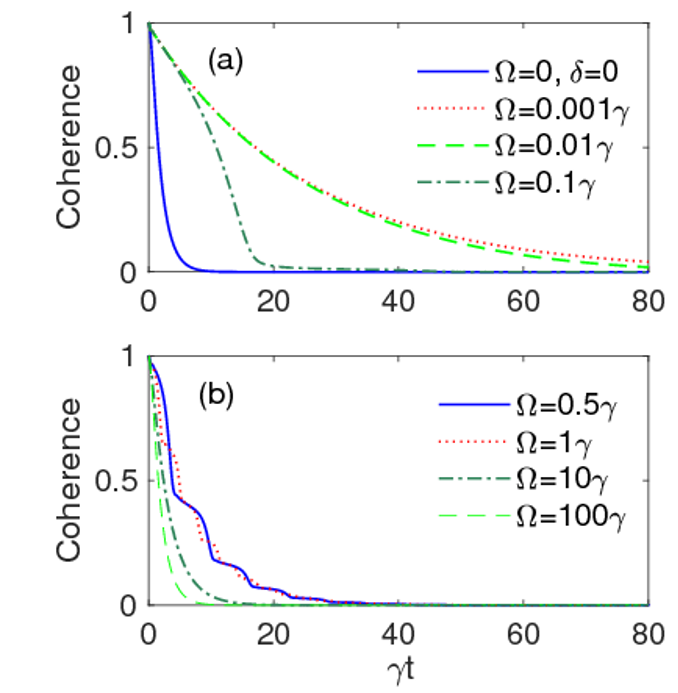}
\end{center}
\caption{Qubit coherence $\zeta(t)$ as a function of scaled time $\gamma t$ for different values of the modulation frequency $\Omega$. The values of other parameters are: $\delta=10\gamma$, $\lambda=3\gamma$ (weak coupling), $\alpha =\beta =1/\sqrt{2}$. Solid-blue line in panel (a) corresponds to the situation in which frequency modulation is off.}
\label{AR2}
\end{figure}

\subsection{Non-Markovianity}
To discuss the non-Markovian character of our system we employ, among the various quantifiers, the well-known measure based on the dynamics of the trace distance between two initially different states $\rho _{1} (0)$ and $\rho _{2} (0)$ of the qubit, identifying information backflows. It is defined as \cite{ref45}
\begin{equation} 
\label{eq:21}
\mathcal{N}=\mathop{\max }\limits_{\rho _{1} (0),\rho _{2} (0)} \int _{\sigma >0}\sigma [t,\rho _{1} (0),\rho _{2} (0)]\mathrm{d}t,  
\end{equation} 
where $\sigma [t,\rho _{1} (0),\rho _{2} (0)]=\mathrm{d}D[\rho _{1} (t),\rho _{2} (t)]/\mathrm{d}t$, with $D[\rho _{1} (t),\rho _{2} (t)]=(1/2)\mathrm{Tr}\left|\rho _{1} (t)-\rho _{2} (t)\right|$ being the trace distance ($\left|X\right|=\sqrt{X^{\dagger} X}$). In Eq.~(\ref{eq:21}) the integration is taken over time intervals when $\sigma >0$ and the maximization is made over all the possible pairs of initial states $\rho _{1} (0)$ and $\rho _{2} (0)$. It is noteworthy that the trace distance is related to the distinguishability between quantum states, whereas its time derivative ($\sigma$) means a flow of information between the system and its environment.
While Markovian processes satisfy $\sigma \leq 0$ for all pairs of initial states $\rho _{1,2} (0)$ at any time $t$, non-Markovian ones admit at least a pair of initial states such that $\sigma >0$ for some time intervals, so that the information flows from the environment back to the system \cite{ref45}. 

For our system, which experiences a dissipative dynamics describable as an amplitude damping channel, the evolved qubit density matrix has the form of Eq.~(\ref{eq:17}). Under this condition, it is known that the non-Markovianity measure $\mathcal{N}$ is maximized for the choice of the initial orthogonal states $(\ket{e}\pm\ket{g})/\sqrt{2}$ of the qubit \cite{ref52} and assumes the expression
\begin{equation}\label{nonMarkMeasure}
\mathcal{N} = - (1/2) \int_{\Gamma<0} \Gamma(t) |C_e(t)| \mathrm{d}t,
\end{equation}
where $\Gamma(t)$ is the effective time-dependent decay rate defined after Eq.~(\ref{eq:18}).

\subsection{Time evolution of single-qubit resources}

We start the quantitative analysis by studying the time evolution of coherence $\zeta(t)$ under a weak coupling regime ($\lambda=3\gamma$), plotted in Fig.~\ref{AR2} for different values of the modulation frequency $\Omega$ and fixed modulation amplitude $\delta=10\gamma$. This evolution is compared to the case when the external driving is off ($\delta=0$, $\Omega =0$), for which coherence disappears at time $t^\ast \sim 10/\gamma$.  
As shown, under the given condition for $\delta$, coherence survives for times longer than $t^\ast$ when $\Omega$ is smaller. Presence of oscillations in the dynamics (see Fig.~\ref{AR2}(b)) indicates the manifestation of non-Markovian effects under weak coupling regime \cite{poggi2017}. However, frequency modulation can even produce negative effects for large values of the modulation frequency, $\Omega\geq 100\gamma$, which are such that the coherence disappears at times $t<t^\ast$.

\begin{figure}[t!]
\begin{center}
\includegraphics[scale=.24]{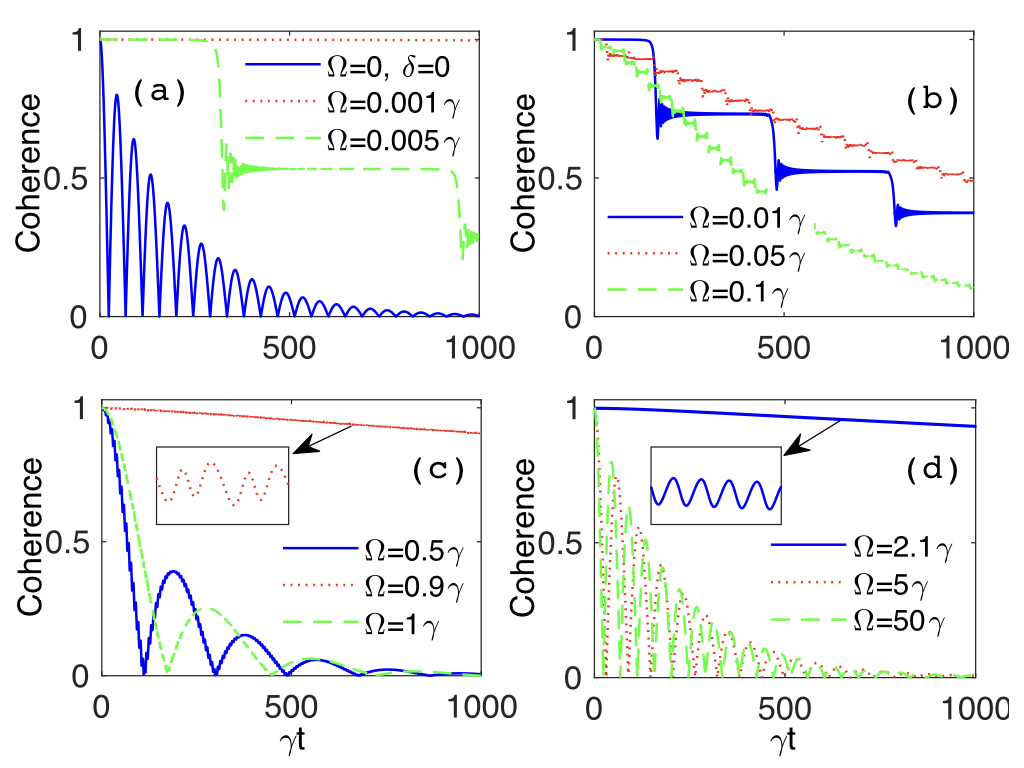}
\end{center}
\caption{Qubit coherence $\zeta(t)$ as a function of scaled time $\gamma t$ for different values of the modulation frequency, for a fixed modulation amplitude $\delta=5\gamma$. Other parameters are: $\lambda=0.01\gamma$ (strong coupling), $\alpha =\beta =1/\sqrt{2}$. Solid-blue line in panel (a) corresponds to the situation in which frequency modulation is off.}
\label{AR7}
\end{figure} 

This fact suggests to see what happens under a strong coupling regime. Fig.~\ref{AR7} shows the time behavior of coherence under such a condition ($\lambda=0.01\gamma$) for different values of modulation frequency at a fixed modulation amplitude 
$\delta=5\gamma$. Interesting results are obtained for $\Omega=0.001\gamma$, 
$\Omega=0.9\gamma$ and $\Omega=2.1\gamma$. The coherence dynamic for 
$\Omega=0.9\gamma$ and $\Omega=2.1\gamma$ is accompanied with rapid and small oscillations, whereas the latter are observed only in the range $\gamma t<400$ for 
$\Omega=0.001\gamma$. This implies that non-Markovianity, associated to backflows of information, should be greater for the first two values of $\Omega$. In general, a non-monotonic behavior of non-Markovianity is expected as a function of $\Omega$, since the intensity of oscillations appears to be very sensitive to different values of the modulation frequency (this aspect shall be treated below). We then point out that, although the plot for $\Omega=0.001\gamma$ seems to perform better than those for $\Omega=0.9\gamma$ and $\Omega=2.1\gamma$ within the time range shown in the figures, our calculations instead show that the coherence lasts much longer for these latter values of the modulation frequency. In general, however, reducing $\Omega$ to values smaller and smaller, with a fixed nonzero $\delta$, helps to prolonging the time when coherence vanishes. This happens because the final effect is to simply detune the central frequency of the high-$Q$ cavity Lorentzian spectral density from the qubit transition frequency \cite{poggi2017,bellomoPhysScr}, as can be deduced from $H_\mathrm{eff}$ of Eq.~(\ref{eq:5}) for $\Omega\rightarrow0$. Nevertheless, this extreme condition is not relevant to our purpose, since we are interested in the modulation of the qubit frequency during the evolution \cite{ref93} and not to a detuning effect.

\begin{figure}[t!]
\begin{center}
\includegraphics[scale=.50]{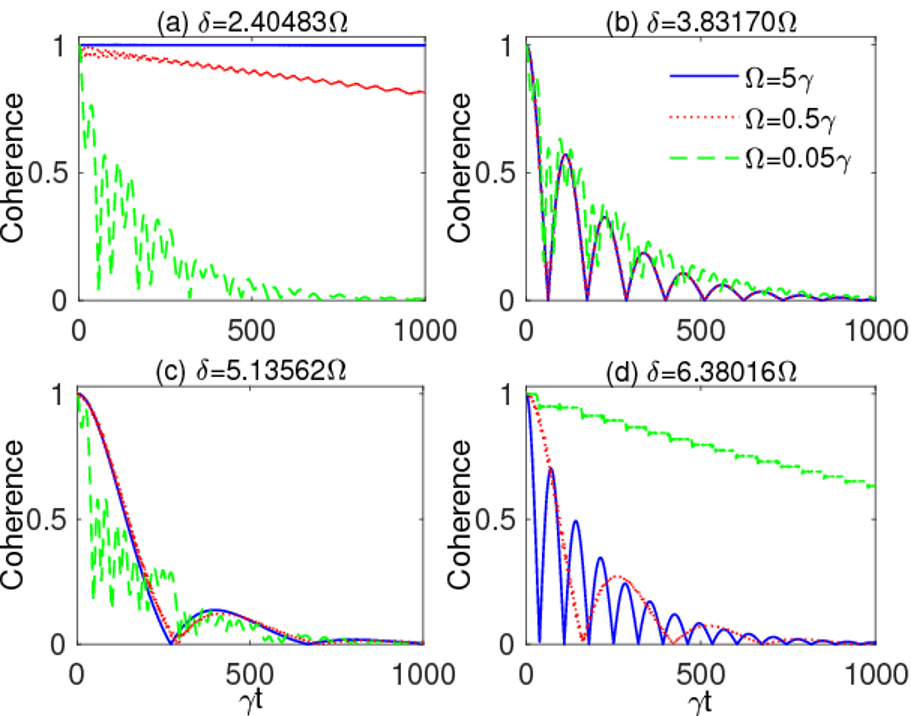}
\end{center}
\caption{Qubit coherence $\zeta(t)$ as a function of scaled time $\gamma t$ for different values of the modulation frequency: $\Omega=0.05\gamma$ (dashed green line), $\Omega=0.5\gamma$ (dotted red line), $\Omega=5\gamma$ (solid blue line). The panels correspond to various values of the modulation amplitudes: (a) $\delta =2.40483\Omega$ $[J_{0}(2.40483\Omega)=0]$, (b) $\delta=3.83170\Omega$ $[J_{1}(3.83170\Omega)=0]$, (c) $\delta=5.13562\Omega$ $[J_{2}(5.13562\Omega)=0]$, (d) $\delta=6.38016\Omega$ $[J_{3}(6.38016\Omega)=0]$. Other parameters are: $\lambda=0.01\gamma$ (strong coupling), $\alpha =\beta =1/\sqrt{2}$.}
\label{AR5}
\end{figure}

Therefore, to gain a more comprehensive understanding of the system evolution, we now study the interplay of the two driving parameters $\delta$ and $\Omega$, which is expected to play a significant role in affecting the decoherence process \cite{agarwalPRA}. Fig.~\ref{AR5} displays the dynamics of coherence, under strong coupling ($\lambda=0.01\gamma$), when the ratio $\delta/\Omega$ is tuned such as to assume values which make the $n$-th Bessel function $J_n$ of Eq.~(\ref{eq:6}) vanish. As seen in Fig.~\ref{AR5}(a), the initial coherence can be strongly protected against the noise by increasing the modulation frequency when $\delta/\Omega$ is such that $J_0$ vanishes. However, this behavior is not general for larger values of the ratio $\delta/\Omega$, as shown in Fig.~\ref{AR5}(b)-(d). The interplay between $\delta$ and $\Omega$ is thus non-trivial, since it is not sufficient to fix 
$\delta/\Omega$ as a zero of the Bessel functions to have long-lasting coherence. In particular, we find that no other settings of $\delta/\Omega$ and $\Omega$ supply a result superseding that occurring for $\delta =2.40483\Omega$ and $\Omega=5\gamma$ (solid blue line of Fig.~\ref{AR5}(a)).

\begin{figure}[t!]
\begin{center}
\includegraphics[scale=.6]{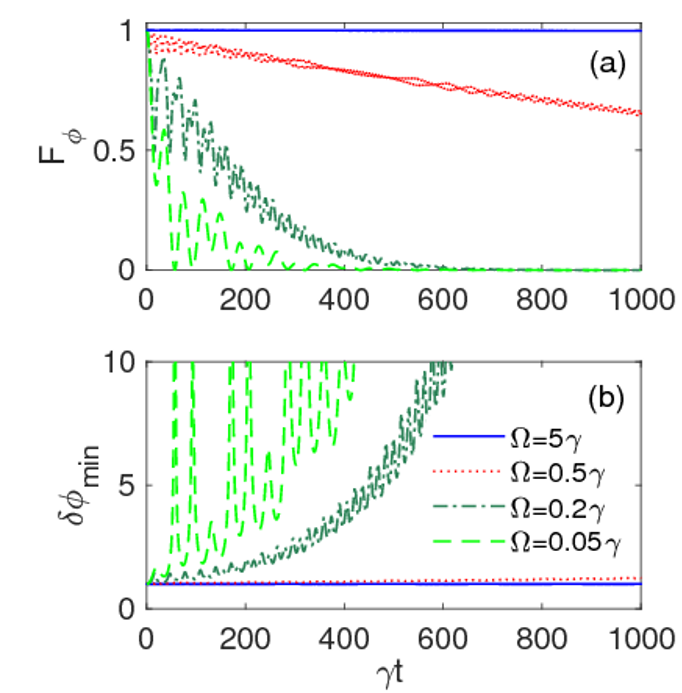}
\end{center}
\caption{(a) Quantum Fisher information $F_{\phi}(t)$ and (b) optimal phase estimation $\delta\phi_\mathrm{min}(t)$ as a function of scaled time $\gamma t$ for various values of modulation frequency: $\Omega=0.05\gamma$ (dashed light green line), $\Omega=0.2\gamma$ (dash-dotted dark green line), $\Omega=0.5\gamma$ (dotted red line), $\Omega=5\gamma$ (solid blue line). Modulation amplitude is fixed at $\delta =2.40483\Omega $. Other parameters are: $\lambda=0.01\gamma$ (strong coupling), $\alpha =\beta =1/\sqrt{2}$.}
\label{AR8}
\end{figure}

We then examine the effect of frequency modulation on the dynamics of both QFI $F_{\phi}(t)$ and optimal phase estimation $\delta\phi_\mathrm{min}(t)$ for the ratio $\delta/\Omega=2.40483$ under strong coupling ($\lambda=0.01\gamma$). From Fig.~\ref{AR8} it is immediately seen, as expected, that maintenance of QFI and a significant improvement in phase estimation can be reached during the evolution by increasing $\Omega$ up to the value $\Omega=5\gamma$.

\begin{figure}[t!]
\begin{center}
\includegraphics[scale=.53]{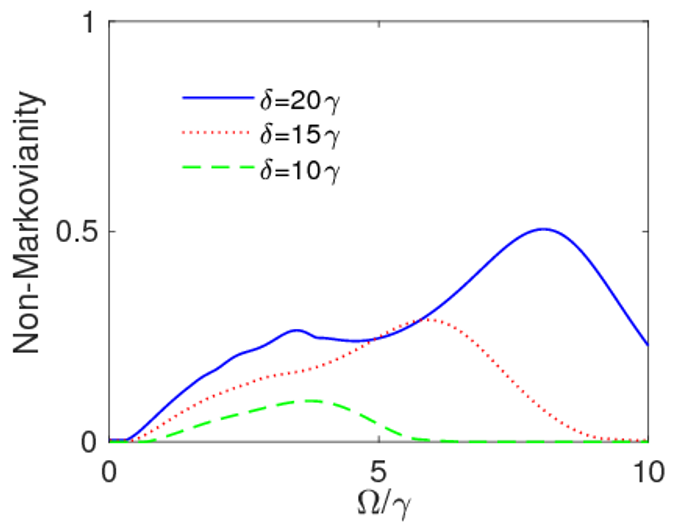}
\end{center}
\caption{Non-Markovianity $\mathcal{N}$ as a function of $\Omega/\gamma$. Values of other parameters are: $\delta=10\gamma$, $\lambda=3\gamma$ (weak coupling), $\alpha =\beta =1/\sqrt{2}$.}
\label{AR3}
\end{figure}

\begin{figure}[t!]
\begin{center}
\includegraphics[scale=.54]{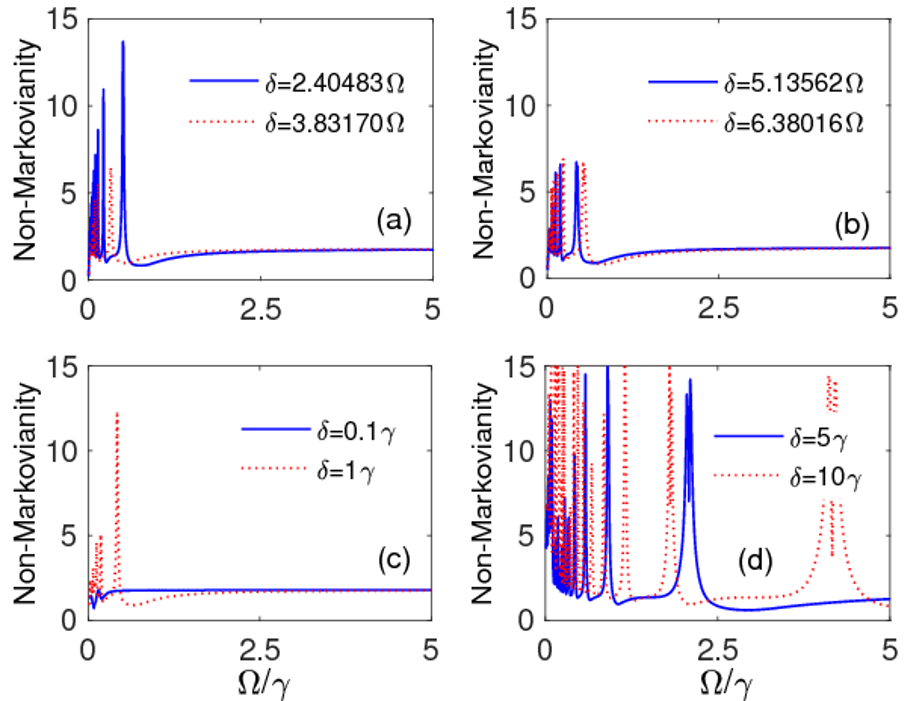}
\end{center}
\caption{Non-Markovianity $\mathcal{N}$ as a function of $\Omega /\gamma $ for: 
(a) $\delta =2.40483\Omega$ $(J_{0}(2.40483\Omega)=0)$ (solid-blue line), 
$\delta=3.83170\Omega$ $(J_{1}(3.83170\Omega)=0)$ (dotted-red line); 
(b) $\delta=5.13562\Omega$ $(J_{2}(5.13562\Omega)=0)$ (solid-blue line); 
(c) $\delta=0.1\Omega$ (solid-blue line), $\delta=1\Omega$ (dotted-red line); 
(d) $\delta=5\Omega$ (solid-blue line), $\delta=10\Omega$ (dotted-red line). 
The values of other parameters are: $\lambda=0.01\gamma$ (strong coupling), $\alpha =\beta =1/\sqrt{2}$.} 
\label{AR4}   
\end{figure}

At this stage, one may ask whether the high coherence protection found above is ultimately linked to strong memory effects. The answer requires the knowledge of the degree of non-Markovianity $\mathcal{N}$ as a function of $\Omega$ under different parameter conditions. This is first done under the weak coupling regime ($\lambda=3\gamma$) for some modulation amplitudes $\delta$ in Fig.~\ref{AR3}. These plots disclose that the frequency modulation process can induce non-Markovian features even in the weak-coupling regime \cite{poggi2017}, which justifies the existence of oscillations in the time evolution of  coherence for $\Omega= 0.5\gamma$ and $\Omega= 1\gamma$ (see Fig.~\ref{AR2}(b)). Moreover, this induced non-Markovianity can be reinforced by increasing the modulation amplitude, while no memory effect is observed for small values of $\Omega$ tending to turn off the frequency modulation. 
In Fig.~\ref{AR4} we then display the behavior of non-Markovianity $\mathcal{N}$ as a function of $\Omega$ under the strong coupling regime ($\lambda=0.01\gamma$). Panels (a) and (b) of this figure consider the cases when the ratio $\delta/\Omega$ corresponds to a zero of the Bessel functions, as in Fig.~\ref{AR5}: one sees that non-Markovian characteristics can be amplified only for $\Omega<1\gamma$, with maximum amplification occurring when $\delta/\Omega$ is the zero of the Bessel function $J_0$. Panels (c) and (d) of Fig.~\ref{AR4} instead consider fixed modulation amplitudes $\delta$ as $\Omega$ changes: it is evident that an increase of $\delta$ enlarges the range of $\Omega$ where non-Markovianity can be enriched, albeit this is achieved only when $\Omega<\delta$. 
As a general behavior, all plots of Fig.~\ref{AR4} show that $\mathcal{N}$ tends to reach approximately the same value as $\Omega/\gamma$ increases, independently of the values of the modulation amplitude $\delta$. Thus, on the basis of our above analysis of qubit dynamics, a more efficient preservation of coherence cannot be related to a higher degree of non-Markovianity (stronger memory effects). 

\begin{figure}[t!]
\begin{center}
\includegraphics[scale=.52]{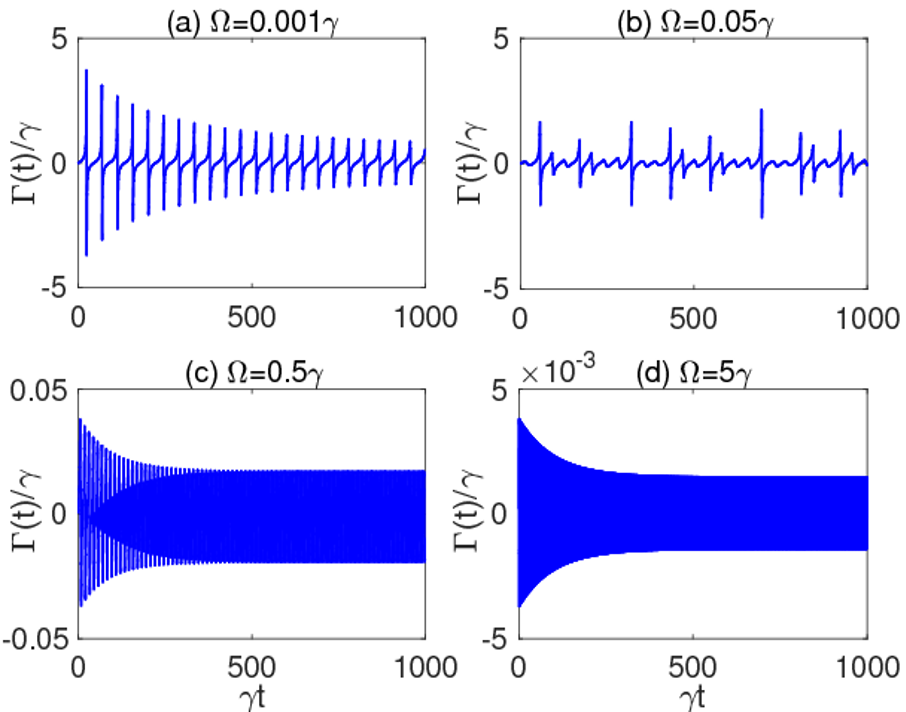}
\end{center}
\caption{Decay rate $\Gamma(t)$ as a function of scaled time $\gamma t$ for different values of the modulation frequency: (a) $\Omega=0.001\gamma$, (b) 
$\Omega=0.05\gamma$, (c) $\Omega=0.5\gamma$, and (d) $\Omega=5\gamma$. Other parameters are those of Fig.~3(a), that is: $\delta =2.40483\Omega$, 
$\lambda=0.01\gamma$ (strong coupling), $\alpha =\beta =1/\sqrt{2}$.}
\label{AR6}
\end{figure}

To have a deeper physical interpretation behind the effective coherence protection, especially exhibited in Fig.~\ref{AR5}(a), we plot in Fig.~\ref{AR6} the time-dependent decay rate $\Gamma(t)$ (in units of $\gamma$), appearing in the qubit master equation of Eq.~(\ref{eq:18}), for some relevant values of $\Omega$. From these plots (see, in particular, panel (d)), it is clear that the main cause of long-lasting coherence is the inhibition of the effective decay rate $\Gamma(t)$ of the qubit. 

Summarizing, according to our analysis, the best performance of coherence preservation controlled by qubit frequency modulation occurs under strong coupling and for values of modulation amplitude $\delta$ and modulation frequency $\Omega$ given, respectively, by
\begin{equation}\label{optpar}
\delta =2.40483\Omega, \quad \Omega=5\gamma,   
\end{equation}
which are expressed in units of the spontaneous decay rate of the qubit $\gamma=\tau_\mathrm{q}^{-1}$. 

\subsection{Experimental context} \label{expcont}

We discuss here how the previous results may impact in experimental contexts where quantum computing platforms allow for accurate preparation of initial states and control of the qubit. This is particularly the case of superconducting qubits, which have been imposing as the building blocks of quantum computer prototypes based on architectures of circuit quantum electrodynamics (circuit-QED) \cite{NatNews2017b,NatNews2017}. 

In order to give a realistic account of our theoretical results, we choose a superconducting transmon qubit embedded in a coplanar resonator. A typical transition frequency for such a qubit is $\omega_0 \approx 50$ GHz, with a spontaneous (free space) decay rate $\gamma \approx 10$ MHz and size $l \approx 200$ $\mu$m \cite{Paik2011PRL}. Within the strong coupling regime, which is the condition of major interest in our study, the considered range of amplitude and frequency modulation parameters is $0\leq \delta \leq 10\gamma$ and $0\leq \Omega \leq 50\gamma$, that guarantee the adiabatic (stationary) regime for the interaction Hamiltonian occurring when $\delta/\omega_0,\Omega/\omega_0\ll1$ \cite{celeriJPB,PhysRevA.57.4784}. Notice that this adiabatic regime appears to be already experimentally achieved for a transmon qubit system in a coplanar resonator \cite{ref97}. These values also imply $\omega_0 \pm \delta \approx \omega_0$ and $\mathbf{k}\cdot\mathbf{r}=(2\pi/\lambda_\mathrm{e})l\approx (\omega_0/c) l\ll 1$, where $\lambda_\mathrm{e}$ is the wavelength of the photon emitted by the qubit and $c$ the speed of light. The latter conditions justify the dipole and rotating wave approximations for the system Hamiltonian along the analysis. Moreover, it is known \cite{poggi2017} that the effects due to a condition of non-rotating wave approximation are negligible when $\omega_0\gg\lambda$, $\lambda$ being the width of the spectral density $J(\omega)$, which is largely satisfied in our study.

From our dynamical results reported above, we know that the coherence of a qubit in a high-$Q$ cavity would naturally (i.e., without frequency modulation) disappear completely at a time 
$t_\mathrm{wc}^\ast \approx 10 \gamma^{-1} = 10 \tau_\mathrm{q}$ under weak coupling (see Fig.~\ref{AR2}(a)) and at $t_\mathrm{sc}^\ast \approx 10^3 \tau_\mathrm{q}$ (after oscillations) under strong coupling (see Fig.~\ref{AR7}(a)). These times are meant as lifetimes of coherent superpositions. For the typical superconducting transmon qubits here considered, which have (free space) relaxation times $\tau_\mathrm{q}\sim 100$ ns \cite{Paik2011PRL}, one would obtain $t_\mathrm{sc}^\ast \approx 100$ $\mu$s, that is comparable to the current  performance achieved in the IBM-Q quantum processor, where a single qubit has average coherence lifetimes of 90 $\mu$s \cite{IBM-Q}. 

Let us now see to which extent our results with frequency-modulated (FM) qubit are able to prolong coherence lifetimes.
Looking at the time behavior obtained under strong coupling with the optimal driving parameters of Eq.~(\ref{optpar}), as displayed by the solid (top) curve of Fig.~\ref{AR5}(a), our calculations estimate the dimensionless time when coherence vanishes at $\gamma t_\mathrm{FM}^\ast \sim 10^7$. This implies a coherence lifetime $t_\mathrm{FM}^\ast \sim 10^7\tau_\mathrm{q}$ and thus $t_\mathrm{FM}^\ast \sim 10^4 \tau_\mathrm{sc}^\ast \approx 1$ s, producing an extension of four orders of magnitude compared to the case of uncontrolled qubit. 
This achievement seems feasible with the current technology. In fact, quality factors of cavities in circuit quantum electrodynamics can be adjusted to high values such that the cavity spectral bandwidth (or photon decay rate) $\lambda$ is smaller than the spontaneous emission decay rate $\gamma$, thus entering the strong coupling regime $\lambda<\gamma$ \cite{Paik2011PRL}. Moreover, frequency modulation of individual superconducting qubits is already realized 
\cite{ref93,ref97}, with the possibility to suitably fix modulation amplitude and frequency to the desired optimal values.

\section{Two-qubit system }\label{secDM}
After individuating the optimal parameters of the driving external field such as to efficiently shield single-qubit coherence from the detrimental effects of the environment, we now extend our study to a composite system of two separated noninteracting qubit-cavity subsystems, namely A and B. Each qubit-cavity subsystem is structured like that considered in the above Sec.~\ref{secBell}, with the transition frequency of each qubit individually modulated. 

\subsection{Initial states and evolved density matrix}
The qubits are initially prepared in the extended Werner-like (EWL) states \cite{ref55}  
\begin{equation} 
\label{eq:26} 
\rho_{\Psi } (0)=r{\left| \Psi  \right\rangle} {\left\langle \Psi  \right|} +\frac{1-r}{4} I, 
\,\
\rho_{\Phi } (0)=r{\left| \Phi  \right\rangle} {\left\langle \Phi  \right|} +\frac{1-r}{4} I, 
\end{equation} 
where $r$ is a measure of the state purity $P$, being for both states $P=\mathrm{Tr}(\rho^2)=(1+3r^2)/4$, $I$ is the $4\times4$ identity matrix and
\begin{equation}
\label{eq:27}  
{\left| \Psi  \right\rangle} =\mu {\left| e_\mathrm{A} e_\mathrm{B}  \right\rangle} +\nu {\left| g_\mathrm{A} g_\mathrm{B}  \right\rangle}, \quad
{\left| \Phi  \right\rangle} =\mu {\left| e_\mathrm{A} g_\mathrm{B}  \right\rangle} +\nu {\left| g_\mathrm{A} e_\mathrm{B}  \right\rangle},
\end{equation}
represent the pure part of the state in the form of Bell-like states with $|\mu|^{2} +|\nu|^{2} =1$. This class of states is quite general, including both Bell states and Werner states. 

The two subsystems, being separated, evolve independently and are individually governed by the Hamiltonain of Eq.~(\ref{eq:1}), so that the total Hamiltonian is simply $H_\mathrm{AB}= H_\mathrm{A} + H_\mathrm{B}$. For the EWL initial states above, the density matrix of the two qubits at time $t$, in the standard (tensor product) computational basis $\{ {\left| 1 \right\rangle} \equiv {\left| e_{A} e_{B}  \right\rangle}, \left| 2 \right\rangle \equiv {\left| e_{A} g_{B}  \right\rangle}, {\left| 3 \right\rangle} \equiv {\left| g_{A} e_{B}  \right\rangle}, {\left| 4 \right\rangle} \equiv {\left| g_{A} g_{B} \right\rangle}\}$, takes the X-type structure \cite{ref55,ref82}
\begin{equation} 
\label{eq:28}  
\rho (t)=
\left(
\begin{array}{cccc} 
{\rho _{11} (t)} & {0} & {0} & {\rho _{14} (t)} \\ 
{0} & {\rho _{22} (t)} & {\rho _{23} (t)} & {0} \\ 
{0} & {\rho _{32} (t)} & {\rho _{33} (t)} & {0} \\
{\rho _{41} (t)} & {0} & {0} & {\rho _{44} (t)}\\
\end{array}
\right), 
\end{equation}
where $\rho_{ji}^{*} (t)=\rho_{ij} (t)$. The elements of this evolved density matrix can be straightforwardly calculated by the well-known method introduced in Ref.~\cite{ref54} for separated qubits, based on the knowledge of the single-qubit density matrix. 
We are interested in the dynamics of quantum resources such as entanglement, discord and coherence associated to the evolved two-qubit state $\rho (t)$. In the following we recall the definitions of their quantifiers. 

\subsection{Quantification of quantum resources}
Entanglement between subsystems of any bipartite quantum system can be measured by concurrence which, for the density matrix given by Eq.(~\ref{eq:28}), is  \cite{lofrancoreview}
\begin{equation}
\label{29} 
\mathcal{C}(t)=2\max \{ 0,\Lambda _{1} (t),\Lambda _{2} (t)\}, 
\end{equation}
where $\Lambda _{1} (t)=|\rho _{14} (t)|-\sqrt{\rho _{22} (t)\rho _{33} (t)} \} $ and $\Lambda _{2} (t)=|\rho _{23} (t)|-\sqrt{\rho _{11} (t)\rho _{44} (t)}$. The two initial EWL states of Eq.~(\ref{eq:26}) have the same concurrence 
$\mathcal{C}(0)=2\max\{0,(|\mu\nu|+1/4)r-1/4\}$ and are thus entangled for $r>1/(1+4|\mu\nu|)$. 

Discord captures all nonclassical correlations between two qubits beyond entanglement \cite{ref18,ref19,ModiRMP}. For a X-structure density matrix, like that of Eq.~(\ref{eq:28}), the analytic expression of discord can be written as \cite{AliPRA2010}
\begin{equation} 
\label{30} 
\mathcal{D}(t)=\min \{\mathcal{D}_{1}(t), \mathcal{D}_{2}(t)\}, 
\end{equation}
where $\mathcal{D}_{j}(t) =h(\rho _{11} +\rho _{33} )+\sum _{i=1}^{4}\lambda _{i} \log _{2}  \lambda _{i}+d_{j}$ ($j=1,2$), with $\lambda _{i}$ being the eigenvalues of the density matrix $\rho (t)$, $h(x)=-x \log _{2} x -(1-x )\log _{2} (1-x )$, $d_{1}=h(\tau )$ with $\tau =(1+\sqrt{(\rho _{11} +\rho _{22} -\rho _{33} -\rho _{44} )^{2} +4(\left|\rho _{14} \right|+\left|\rho _{23} \right|)^{2} } )/2$, and $d_{2} = -h(\rho _{11} +\rho _{33}) -\sum _{i=1}^{4}\rho _{ii} \log _{2}  \rho _{ii}$ (the explicit time-dependence of the density matrix elements has been omitted for simplicity). The two initial EWL states have the same discord $\mathcal{D}(0)$ as a function of $r$ and $\mu$ whose explicit expression, obtainable by the previous formulas, is not reported here.

Concerning coherence, the definition given in Eq.~(\ref{eq:19}) is immediately applicable to the two-qubit evolved density matrix of Eq.~(\ref{eq:28}), so to give the two-qubit coherence with respect to the computational basis
\begin{equation}
\zeta_2(t)=|\rho_{14}(t)|+|\rho_{41}(t)|+|\rho_{23}(t)|+|\rho_{32}(t)|.
\end{equation}
Both the initial EWL states of Eq.~(\ref{eq:26}) have the same coherence 
$\zeta_2(0)=2r|\mu\nu|$ and are thus coherent for any $r>0$ provided that $\mu\neq 0$. Notice that for $\mu=\nu=1/\sqrt{2}$, that is the pure part of the EWL states is a (maximally entangled) Bell state, one has $\zeta_2(0)=r$. 

We are now ready to display the dynamics of the relevant quantities defined above.

\subsection{Time evolution of two-qubit resources}

\begin{figure}[t!]
\begin{center}
\includegraphics[scale=.25]{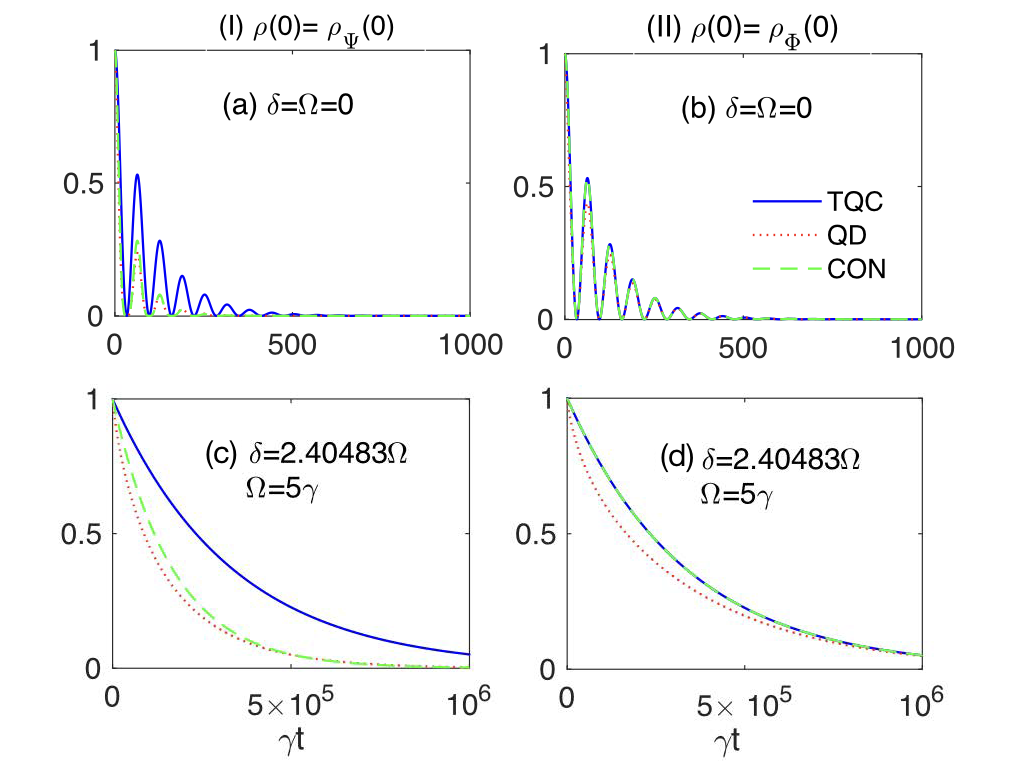}
\end{center}
\caption{Time evolution of two-qubit coherence (TQC) $\zeta_2(t)$ (solid blue line), quantum discord (QD) $\mathcal{D}(t)$ (dotted red line) and concurrence (CON) $\mathcal{C}(t)$ (dashed green line) for the initial states $\rho_{\Psi}(0)$ (column I) and $\rho_{\Phi}(0)$ (column II), with $r=1$ and $\mu=\nu=1/\sqrt{2}$ (Bell states), under $\lambda=0.01\gamma$ (strong coupling). Panels (a), (b): uncontrolled dynamics (no frequency modulation). Panels (c), (d): controlled dynamics (individual frequency modulation) with optimal parameters $\delta=2.40483\Omega$, $\Omega=5\gamma$.}
\label{AR9plus}
\end{figure}

We assume that the subsystems are identical, that is characterized by the same values of qubit-environment parameters, with the optimal frequency modulation parameters fixed as in Eq.~(\ref{optpar}), under the strong coupling regime with $\lambda=0.01\gamma$. Such a choice has the advantage to make us directly focus on the best quantitative protection of two-qubit resources attainable by individual qubit frequency modulation under a given strong coupling condition. To acquire information about the protection efficiency of individual frequency modulation, we report the dynamics of $\mathcal{C}(t)$, $\mathcal{D}(t)$ and $\zeta_2(t)$ starting from the initial EWL states of Eq.~(\ref{eq:26}) with $r=1$ and $\mu=\nu=1/\sqrt{2}$, which reduces them to the Bell states, respectively, $\ket{\Psi} = (\ket{ e_\mathrm{A} e_\mathrm{B}} +\nu \ket{g_\mathrm{A} g_\mathrm{B}})/\sqrt{2}$ and $\ket{\Phi} =(\mu \ket{e_\mathrm{A} g_\mathrm{B}} +\nu \ket{g_\mathrm{A} e_\mathrm{B}})/\sqrt{2}$. Fig.~\ref{AR9plus} immediately shows that individual qubit control by frequency modulation enables lifetimes of quantum resources $\bar{t}$ (that is, the time when they completely disappear) which are three orders of magnitude longer than the lifetimes when modulation is off. In particular, we find $\bar{t}_\mathrm{FM} \sim10^6 \tau_\mathrm{q}$ against $\bar{t} \approx 500\tau_\mathrm{q}$ without frequency modulation. Concerning dynamical details, one sees that while the evolution of coherence is the same for both initial states, discord and entanglement vanish earlier for the initial state $\rho_{\Psi}(0)=\ket{\Psi}\bra{\Psi}$ compared to $\rho_{\Phi}(0)=\ket{\Phi}\bra{\Phi}$. Another difference in the time behavior of quantum resources for the two initial states is that entanglement and coherence coincide at any time for $\rho_{\Phi}(0)=\ket{\Phi}\bra{\Phi}$ (panels (b) and (d) of Fig.~\ref{AR9plus}), whereas this is not the case for $\rho_{\Psi}(0)=\ket{\Psi}\bra{\Psi}$.  

\begin{figure}[t!]
\begin{center}
\includegraphics[scale=.53]{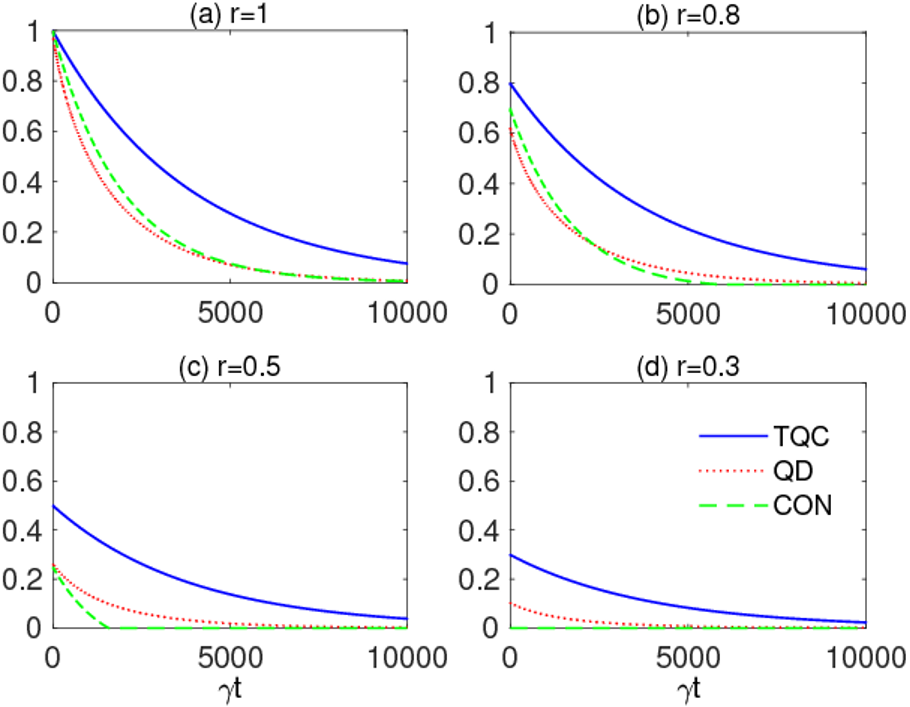}
\end{center}
\caption{Two-qubit coherence (TQC) $\zeta_2(t)$ (solid blue line), quantum discord (QD) $\mathcal{D}(t)$  (dotted red line) and concurrence (CON) $\mathcal{C}(t)$ (dashed green line) for the two-qubit initial state $\rho_{\Psi}(0)$ as functions of scaled time $\gamma t$ for different degrees of purity: (a) $r=1$, (b) $r=0.8$, 
(c) $r=0.5$, and (d) $r=0.3$. Other parameters are: $\mu=\nu=1/\sqrt{2}$, $\delta=2.40483\Omega$, $\Omega=5\gamma$ and $\lambda=0.1\gamma$ (strong coupling).}
\label{AR9}
\end{figure}

\begin{figure}[t!]
\begin{center}
\includegraphics[scale=.24]{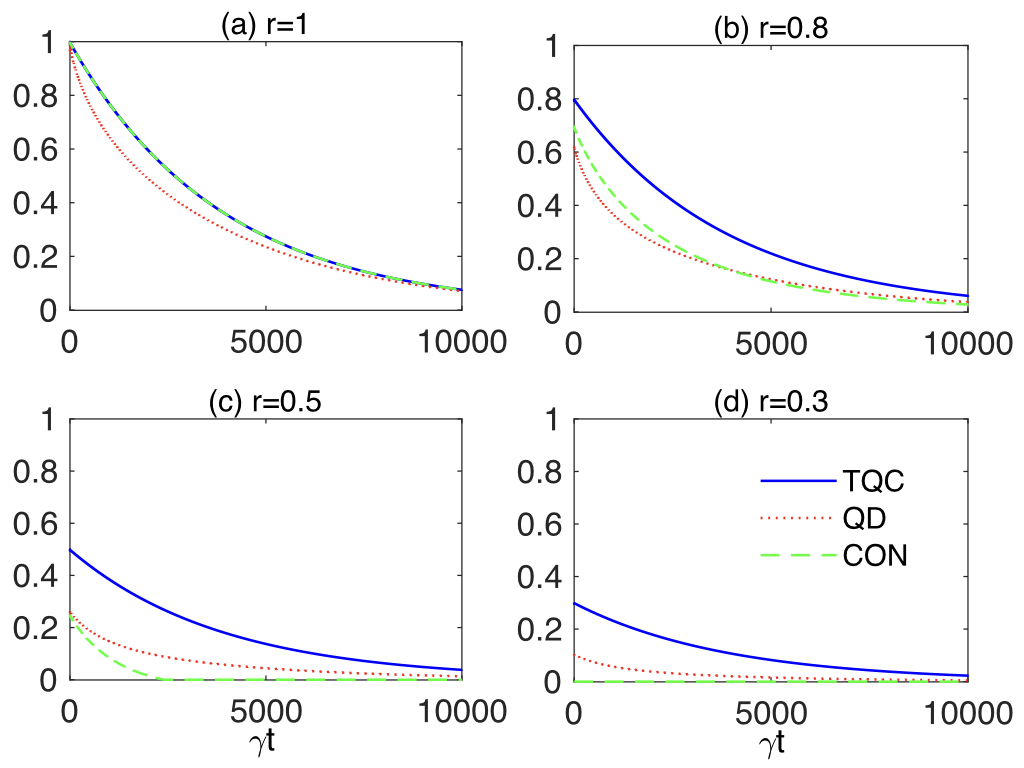}
\end{center}
\caption{Two-qubit coherence (TQC) $\zeta_2(t)$ (solid blue line), quantum discord (QD) $\mathcal{D}(t)$  (dotted red line) and concurrence (CON) $\mathcal{C}(t)$ (dashed green line) for the two-qubit initial state $\rho_{\Phi}(0)$ as functions of scaled time $\gamma t$ for different degrees of purity: (a) $r=1$, (b) $r=0.8$, (c) $r=0.5$, and (d) $r=0.3$. Other parameters are taken as in Fig.~\ref{AR9}.}
\label{AR10}
\end{figure}

Finally, we analyze an attenuated strong coupling condition $\lambda=0.1\gamma$ for decreasing values of the purity parameter $r$ in the EWL states of Eq.~(\ref{eq:26}), in order to take into account both lower quality factors of cavities and possible imperfections in the initial state preparation. From Figs.~\ref{AR9} and \ref{AR10}, as expected, one sees that the amount of all the quantum resources diminishes for smaller $r$ and tends to zero faster. For $r=0.3$ entanglement is always zero, since $\mathcal{C}(0)=0$. We also observe that two-qubit coherence is the more robust resource of the three ones, entanglement being instead the more fragile. Once again, the general emerging result is that all the quantum resources are efficiently shielded from the noise by qubit frequency modulation, their lifetime $\bar{t}$ being extended of about three orders of magnitude with respect to the case without modulation. This is seen, for instance, in the case $r=1$ of Figs.~\ref{AR9}(a) and \ref{AR10}(a) where $\bar{t}_\mathrm{FM} \approx 10^4 \tau_\mathrm{q}$, against $\bar{t} \approx 50\tau_\mathrm{q}$ found in absence of frequency modulation.

These results assume experimental significance in the context of circuit quantum electrodynamics, where the required individual control of the subsystems has been already discussed above (Sec.~\ref{expcont}) and a reliable preparation of entangled states of superconducting qubits has been implemented \cite{Dicarlo2010,Rigetti2012}.
For example, entangled states with purity $\approx 0.87$ and fidelity to ideal Bell states $\approx 0.90$ have been generated in the laboratory by using a two-qubit interaction mediated by a cavity bus in a circuit quantum electrodynamics architecture \cite{Dicarlo2010}. These states may be approximately described as the EWL states of Eq.~(\ref{eq:26}) with $r=r_\mathrm{exp}\approx0.91$ and $\mu=\nu=1/\sqrt{2}$ \cite{ref77}, which is just the configuration of the initial states we have assumed in our study.

\section*{Discussion}
\label{secU}

In this work we have investigated in detail the effect of individual modulation of qubit transition frequency in protecting quantum resources from the detrimental effect of the environment, constituted by leaky high-$Q$ cavities. Compared to previous studies about qubit frequency modulation, our work has the merit (i) to provide an extensive quantitative analysis of the dynamics of quantum resources under a known control technique of qubit systems and (ii) to supply its potential impact on a cutting-edge technology employed for quantum computer prototypes, such as circuit quantum electrodynamics with superconducting qubits.

We have first put our attention on a single frequency-modulated qubit embedded in a cavity, determining optimal modulation parameters $\delta$ (modulation amplitude) and $\Omega$ (modulation frequency) and the qubit-cavity coupling regime allowing a long-time preservation of coherence. We have found that, under a strong coupling regime, qubit coherence lifetimes can be extended of orders of magnitude with respect to the case when modulation is off. In particular, when the ratio between the cavity spectral bandwidth $\lambda$ and the spontaneous emission rate of the qubit $\gamma$ is $\lambda/\gamma = 10^{-2}$ we have shown that this coherence lifetime extension can be of four orders of magnitude. We have also seen that the same conditions guarantee a maintenance of quantum Fisher information with a consequent relevant advantage in phase estimation processes of the single-qubit state during the evolution. We have individuated the inhibition of the effective decay rate of the qubit as the mechanism underlying the efficient dynamical preservation of quantum coherence, which is therefore not due to stronger non-Markovianity (memory effects) of the system. 

We have then exploited the findings of the single-qubit case to control resources like entanglement, discord and coherence in a composite system of two separated qubits. Such a study, so far unexplored, has revealed itself very useful as a first step towards understanding the scalability of the control procedure to individually addressable subsystems. We have found that individual qubit frequency modulation is still capable to increase the lifetimes of the desired quantum resources of orders of magnitude compared to their natural (uncontrolled) disappearance times. In fact, albeit the gain in the two-qubit system is weaker than in the single-qubit case, with the optimal parameters one may reach lifetime prolongations of three orders of magnitude. 

We have discussed the experimental feasibility of the systems, particularly considering the state-of-art achievements in the context of circuit quantum electrodynamics. Setups of superconducting qubits can indeed modulate the qubit transition frequency by external control, reach strong qubit-cavity coupling conditions and permit high-fidelity initial state preparation. The explicit values of the control parameters provided here to achieve the high preservation of quantum resources appear to be feasible and constitute a useful practical information for experimental applications within the so-called noisy intermediate-scale quantum (NISQ) devices \cite{NISQ}. Moreover, the technique is straightforwardly applicable to an array of many separated qubit-cavity subsystems to further investigate scalability. Our results provide novel insights for efficient experimental strategies against decoherence towards reliable quantum-enhanced technologies.



 \end{document}